\documentclass{article}

\usepackage[numbers]{natbib}

\usepackage[preprint]{neurips_2025}

\usepackage[utf8]{inputenc} 
\usepackage[T1]{fontenc}    
\usepackage{hyperref}       
\usepackage{url}           
\usepackage{booktabs}       
\usepackage{amsfonts}      
\usepackage{nicefrac}      
\usepackage{microtype}      
\usepackage{xcolor}         
\usepackage{graphicx}
\usepackage{subcaption}
\usepackage{amsmath} 
\usepackage{pifont} 

\usepackage{colortbl}    
\usepackage{amssymb}   
\usepackage{multirow}
\usepackage{float}
\usepackage{amsmath, amssymb}
\usepackage{algorithm}
\usepackage{algpseudocode}
\usepackage{bm}
\newcommand{\xmark}{\ding{55}}
\title{
SpeechVerifier: Robust Acoustic Fingerprint against Tampering Attacks via Watermarking
}

\author{%
    \textbf{Lingfeng Yao}$^{1}$\thanks{Equal contribution.} \qquad \textbf{Chenpei Huang}$^{1}$\footnotemark[1] \qquad \textbf{Shengyao Wang}$^2$ \qquad \textbf{Junpei Xue}$^2$ \\ \qquad \textbf{Hanqing Guo}$^{3}$ \qquad \textbf{Jiang Liu}$^2$ \qquad \textbf{Xun Chen}$^4$ \qquad \textbf{Miao Pan}$^1$ \\
    $^1$University of Houston \\
    $^2$Waseda University \\
    $^3$University of Hawaii at Mānoa \\
    $^4$Independent Researcher
}

\begin{document}
\maketitle
\begin{abstract}
With the surge of social media, maliciously tampered public speeches, especially those from influential figures, have seriously affected social stability and public trust.
Existing speech tampering detection methods remain insufficient: they either rely on external reference data or fail to be both sensitive to attacks and robust to benign operations, such as compression and resampling.
To tackle these challenges, we introduce SpeechVerifer to proactively verify speech integrity using only the published speech itself, i.e., without requiring any external references.
Inspired by audio fingerprinting and watermarking, SpeechVerifier can (i) effectively detect tampering attacks, (ii) be robust to benign operations and (iii) verify the integrity only based on published speeches. 
Briefly, SpeechVerifier utilizes multiscale feature extraction to capture speech features across different temporal resolutions. Then, it employs contrastive learning to generate fingerprints that can detect modifications at varying granularities. These fingerprints are designed to be robust to benign operations, but exhibit significant changes when malicious tampering occurs. To enable speech verification in a self-contained manner, the generated fingerprints are then embedded into the speech signal by segment-wise watermarking. Without external references, SpeechVerifier can retrieve the fingerprint from the published audio and check it with the embedded watermark to verify the integrity of the speech. Extensive experimental results demonstrate that the proposed SpeechVerifier is effective in detecting tampering attacks and robust to benign operations.
\end{abstract}

\section{Introduction}
Audio serves as an important information carrier that is widely used in news reporting, legal evidence, and public statements. However, the rapid development of audio editing tools~\cite{wang2023audit} and text-to-speech (TTS) generation models~\cite{oord2016wavenet, ping2018deep, wang2017tacotron, huang2023make} has significantly lowered the technical barriers for speech manipulation and synthesis. While these techniques benefit content creation and entertainment, they also enable attackers to tamper speech content with ease. Public speeches and statements, especially made by influential figures, have become prime targets for attacks due to their huge social impact. Tampered speech can cause the spread of misinformation, undermine public trust, and even threaten social stability. Moreover, the prevalence of social media platforms accelerates the circulation of tampered audio, posing challenges to ordinary people in identifying the authenticity from numerous sources. For instance, some statements of U.S. Presidents have been repeatedly edited and manipulated, and broadly redistributed on various social media platforms across the internet~\cite{biden2023,trump2024}. As such doctored speeches reach the public, doubt naturally arises concerning their integrity. Currently, verifying the truth often requires cross-checking information across multiple social media platforms, a process that is time-consuming and prolongs the spread of misinformation. These challenges highlight a critical need: Is it possible to proactively protect publicly shared speech against tampering attacks while still allowing it to be freely stored, distributed, and reshared?

Existing approaches against speech tampering attacks can roughly be categorized into two groups: passive detection and active protection. Passive detection methods~\cite{rodriguez2010audio, yang2008detecting, pan2012detecting, leonzio2023audio, blue2022you} primarily rely on deep binary classifiers trained to identify subtle artifacts introduced by tampering operations. While they show reasonable performance against known attacks, their sensitivity to unseen or sophisticated manipulations remains limited. Moreover, passive detection alone cannot verify whether the speech content originates from the claimed speaker, leaving systems vulnerable to impersonation-based attacks~\cite{khan2022voice}. On the other hand, active protection methods aim to ensure content integrity by embedding auxiliary information into the audio signal or extracting it during verification. Common approaches include cryptographic hashing~\cite{steinebach2003watermarking} and fragile watermarking~\cite{renza2018authenticity}, both of which can reliably detect content alterations. However, these methods are highly sensitive to benign operations such as compression or resampling, restricting their applicability in real-world distribution scenarios. Furthermore, hash-based verification typically requires transmitting or retrieving external reference hash values, limiting their ability for independent self-verification of the published speech audio.

To address the issues above, a desired speech verification design should have the following properties: (1) \textbf{Convenient to use}: the integrity of the speech can easily be verified by the general public without requiring external references. (2) \textbf{Sensitive to tampering attacks}: it can reliably detect any malicious edits, including subtle semantic (e.g., can $\Leftrightarrow$ cannot) or speaker-related (e.g., timbre) changes. (3) \textbf{Robust to benign operations}: it should be robust to typical benign audio operations, especially commercial-off-the-shelf codecs (e.g., AAC or Opus in Instagram/TikTok), ensuring usability in sharing and distribution. Therefore, in this paper, we propose SpeechVerifier, a proactive acoustic fingerprint-based speech verification design that jointly utilizes semantic content and speaker identity. Specifically, SpeechVerifier uses multiscale feature extraction to capture speech features across different temporal resolutions. Then, it employs contrastive learning to generate fingerprints that can detect modifications at varying granularities. These fingerprints are designed to be robust to benign operations, but exhibit significant changes when malicious tampering occurs. To enable speech verification in a self-contained manner, the generated fingerprints are then embedded into the speech signal by segment-wise watermarking. Without a copy of the original authentic speech, SpeechVerifier can retrieve the fingerprint from the published audio and check it with the embedded watermark to verify the integrity of the speech. Our salient contributions are summarized as follows.
\begin{itemize}
    \item We propose SpeechVerifier, a proactive speech verification design against tampering attacks, which enables users to verify speech integrity without accessing original speech recordings.
    \item To enable \textbf{self-contained verification}, we leverage audio watermarking to embed discriminative fingerprints into the speech signal, allowing for verifying the integrity only from the watermarked audio.    
    \item We develop a five-step algorithm that extracts multiscale features and applies contrastive learning to generate binary fingerprints, which are \textbf{robust to benign} operations yet\textbf{ sensitive to malicious} manipulations. 
    \item Extensive experiments across diverse audio manipulation scenarios show that SpeechVerifier is effective in detecting tampering attacks and robust to benign operations.
\end{itemize}

\section{Related works}
\textbf{Detect speech tampering passively based on acoustic features.}
Audio editing process can generate artifacts or modify natural acoustic features within human speech. For example, frame offset~\cite{yang2008detecting}, inconsistent noise~\cite{pan2012detecting}, and even discontinued electric network frequency~\cite{rodriguez2010audio,esquef2014edit}, are identified as evidence of tampering. Using such patterns, ``passive'' detectors can be trained as binary classifiers using labeled clean and tampered audio. However, these methods are less effective when facing deepfake audio. Advanced deepfake techniques can synthesize high-fidelity speech with few or no detectable artifacts, making conventional patterns unreliable. To address this, recent work has explored more subtle acoustic properties, such as fluid dynamics and articulatory phonetics~\cite{blue2022you}. Nevertheless, as the deepfake models evolve, relying solely on passive detection may not be sufficient against future attacks.

\textbf{Protect genuine audio based on integrity verification.}
Proactive defense provides another direction to detect speech tampering. In general, critical information is extracted from the original audio and condensed into auxiliary data (or ``meta'' data). This auxiliary data then serves as a reference for verification: one extracts the same data from the test audio, and if it matches, it indicates that the test audio is free of audio editing, and vice versa. Cryptographic hashing, which transforms the digital audio files into discrete bytes, is one of the proactive defenses~\cite{zakariah2018digital}. However, hashing operations are too sensitive to tolerate common operations from regular users, such as audio compression, resampling, resulting in a high false alarm rate. Another method is fragile watermarking~\cite{renza2018authenticity}, where sensitive watermarks are directly embedded into audio signals and checked for changes. However, this method is also sensitive to minor perturbations, limiting the free and practical distribution of audio. \citet{geproactive} propose a proactive defense approach against speaker identity manipulation, which embeds speaker embeddings into speech using audio watermarking. However, their method focuses only on speaker-identity attacks and cannot detect semantic content alterations. Therefore, existing proactive audio protection methods do not simultaneously achieve robustness against benign operations, sensitivity to malicious tampering, and independence from external verification channels.

{
\setlength{\textfloatsep}{2pt}  
\begin{figure}[!t]
  \centering
  \includegraphics[width=1.0\linewidth]{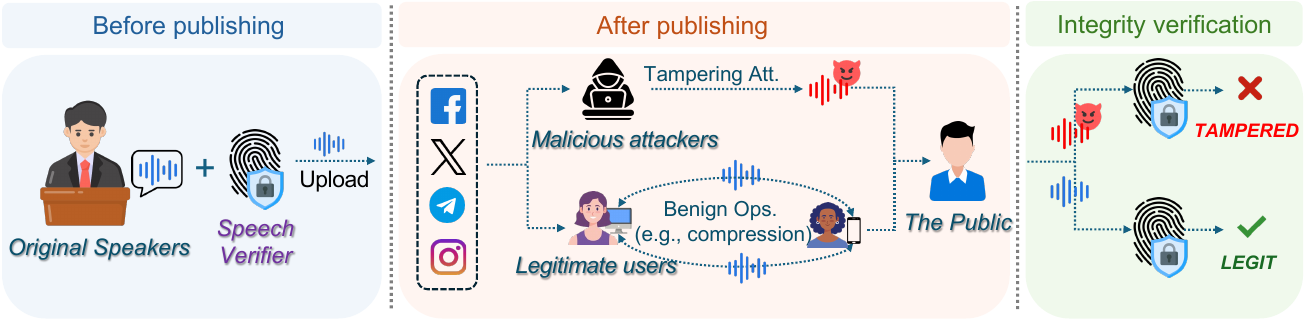}
  \caption{System overview of the proposed SpeechVerifier.} 
  \label{fig:system_overview}
\end{figure}
}

\section{Motivation}
\subsection{Problem Definition}
As shown in Figure~\ref{fig:system_overview}, the scenario considered in our study includes four parties: 1) \textbf{Original speakers}, such as public institutions and celebrities, who publish statements or speeches on social media platforms. 2) \textbf{Legitimate users}, who download and repost these recordings to increase their dissemination. 3) \textbf{Malicious attackers}, which employ audio editing or voice conversion techniques, either changing the original semantic meaning or impersonating the speaker. 4) \textbf{The public}, who are exposed to conflicting audio sources, requiring a practical and reliable method to verify the integrity of a given speech recording.

\subsection{Definition of Malicious Tampering vs. Benign Operations}
We define tampering attacks as malicious audio modifications that alter the semantic content or the speaker identity. Typical malicious operations include audio splicing, deletion, substitution, silencing, text-to-speech (TTS) synthesis, and voice conversion. In contrast, benign operations refer to common audio transformations that occur during legitimate storage, transmission, or distribution processes, such as compression, reencoding, resampling, and noise suppression. These operations do not affect semantic content or speaker identity, and are therefore not considered tampering attacks. To better evaluate the impact of tampering, we further categorize malicious operations into three levels - Minor, Moderate, and Severe - based on their impacts on semantic content and speaker identity. A detailed distinction between malicious and benign audio operations, along with specific examples, is provided in Appendix~\ref{apx:A2}.

\subsection{Limitations of Coarse-Grained Acoustic Feature Based Similarity Comparison} \label{section3.3}
An intuitive approach to speech verification is to check whether the published speech audio sounds like the original one, which is technically to measure acoustic feature similarity. Following this intuition, we compared the similarity distributions of 3 categories: (i) the original audio and the audio modified by benign operations (``Benign''), (ii) the original audio and the audio modified by malicious operations (``Malicious''), and (iii) the original audio and the arbitrarily selected unrelated audio (``Cross''). Moreover, we compare the similarity distributions across malicious operations with varying degrees of tampering, as introduced earlier. Two types of acoustic features were used: Mel-frequency cepstral coefficients (MFCC)~\cite{davis1980comparison} and deep representations from wav2vec2.0~\cite{baevski2020wav2vec}. As shown in Figure~\ref{fig:sim1} and Figure~\ref{fig:sim2}, the similarity distributions based on wav2vec embeddings for ``Benign" and ``Malicious" overlap, and the similarity scores tend to decrease as the extent of tampering increases. This observation indicates that such similarity comparison can only measure the extent of modification, but cannot distinguish the types of modification operations (i.e., benign or malicious ones). Similar results using MFCC features are provided in Appendix~\ref{apx:B3}. These observations demonstrate that coarse-grained acoustic feature based similarity checking tends to ignore minor but semantically important speech edits. For instance, altering the phrase ``do not'' to ``do'' in a 20-second speech affects only about 0.2 seconds, and similarity remains more than 99\%, while there is a huge semantic difference. Moreover, these methods require access to the original authentic audio, which is impractical in many scenarios. These observations highlight two key challenges:

\begin{figure}[t]
    \centering
    \begin{subfigure}[t]{0.32\textwidth}
        \centering
        \includegraphics[width=\textwidth]{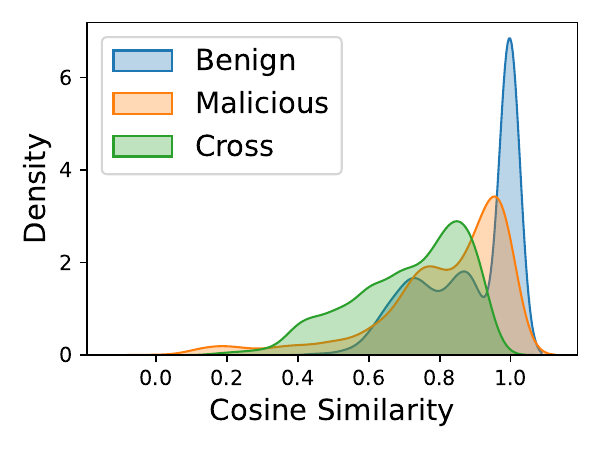}
        \caption{}
        \label{fig:sim1}
    \end{subfigure}
    \hfill
    \begin{subfigure}[t]{0.32\textwidth}
        \centering
        \includegraphics[width=\textwidth]{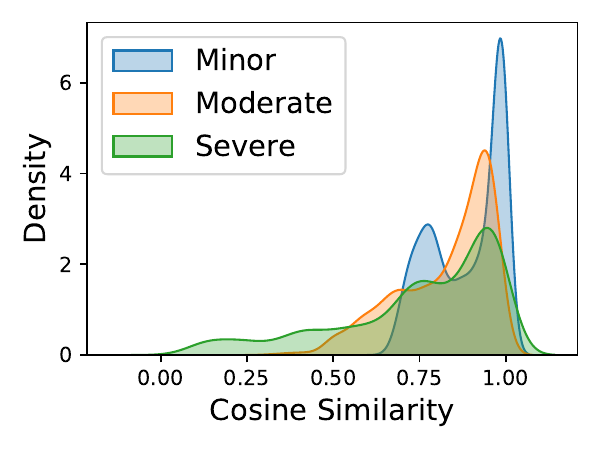} 
        \caption{}
        \label{fig:sim2}
    \end{subfigure}
    \hfill
    \begin{subfigure}[t]{0.32\textwidth}
        \centering
        \includegraphics[width=\textwidth]{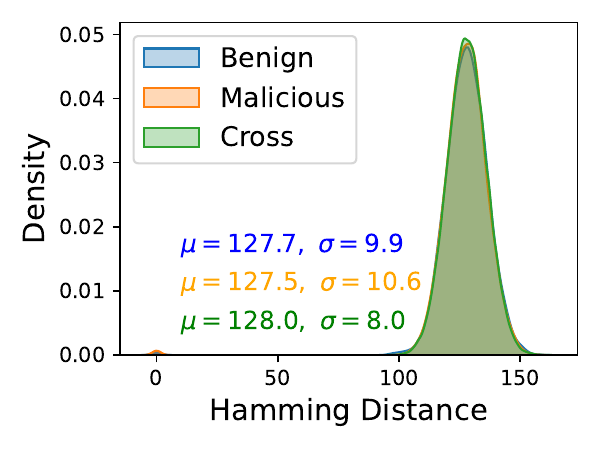} 
        \caption{}
        \label{fig:sha256}
    \end{subfigure}
    \caption{Probability distributions: (a) wav2vec embedding similarity to the original audio under different modifications; (b) wav2vec embedding similarity to the original audio at different tampering levels; (c) SHA256 Hamming distance to the original audio under different modifications.
}
    \label{fig:observation}
\end{figure}

\textbf{Challenge 1: Insufficient sensitivity to semantic tampering attacks.} The coarse-grained acoustic feature based similarity checking methods fail to distinguish benign operations from malicious ones, because they are not sensitive enough to semantic tampering attacks.

\textbf{Challenge 2: Dependence on the original authentic audio.} The similarity comparison methods have to use the original authentic audio as the reference, which is not always applicable in practice.

\vspace{-4mm}
\subsection{Limitations of Cryptographic Hashing Based Methods}
Given Challenge 1, it is worth exploring some other speech verification methods that are sensitive to tampering attacks. A potential candidate is cryptographic hashing~\cite{cryptohash}, which generates a digest of the entire audio file. Due to its high sensitivity, even the slightest modifications can be detected, so it can detect malicious tampering attacks effectively. However, such high sensitivity is not always positive in practice. Hash values may also change significantly under benign operations that do not affect either the semantic content or speaker identity. As shown in Figure~\ref{fig:sha256}, the Hamming distances between original audio samples and their corresponding benign variants, malicious variants, and unrelated audio samples are all similarly large when using cryptographic hashing. This observation indicates that cryptographic hashing cannot distinguish benign operations from malicious tampering attacks. Moreover, hashing-based methods rely on external hashed results beyond the published audio itself for verification, which introduces extra cost. These observations expose two key challenges:

\textbf{Challenge 3: Lack of robustness to benign operations.} Hash values change significantly even under benign operations, making it too sensitive to use for speech verification.

\textbf{Challenge 4: Dependence on external reference hash values.} Hashing based methods rely on external reference hash values for speech verification. That introduces extra overhead and may not be convenient for speech forwarding on the online platforms in practice.

\begin{figure}[h]
  \centering
  \includegraphics[width=1.0\linewidth]{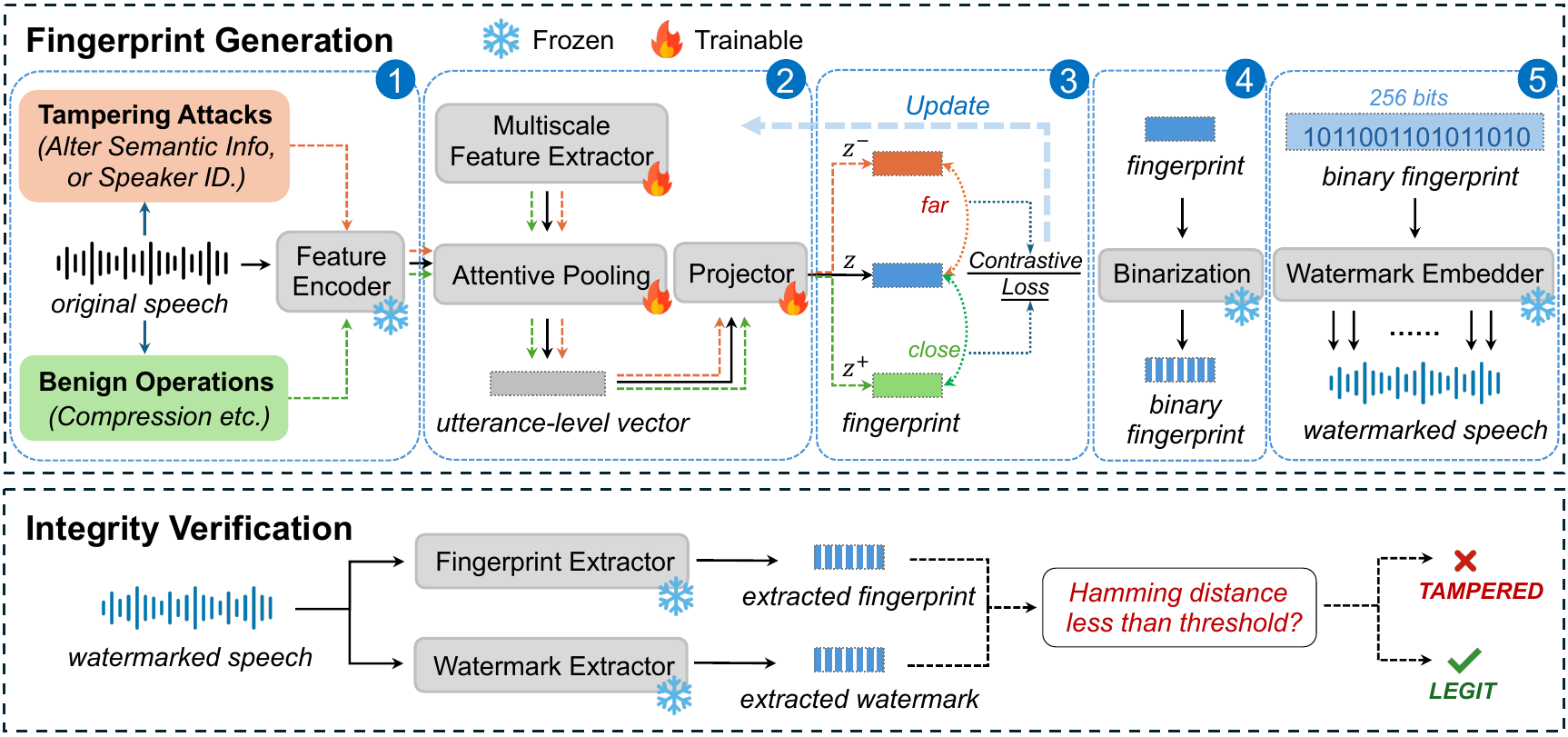}
  \caption{A sketch of the proposed SpeechVerifier design including speech fingerprint generation (top) and integrity verification (bottom).} 
  \label{fig:framework}
\end{figure}

\section{Methodology}
\subsection{SpeechVerifier Design Overview}
To address those challenges, we propose {SpeechVerifier}, a proactive speech integrity verification design, which is (i) sensitive to tampering attacks, (ii) robust to benign operations, and (iii) convenient to use by the public since it verifies the published speech audio's integrity in a self-contained manner. As the sketch shown in Figure~\ref{fig:framework}, SpeechVerifier consists of two stages: fingerprint generation and dual-path integrity verification.

The speech fingerprint generation in SpeechVerifier has five steps:
(1) \textbf{Frame-Level Feature Encoding (Speech to Representation)}: raw speech is encoded into frame-level representations that preserve acoustic information;
(2) \textbf{Multiscale Acoustic Feature Extraction (Representation to Vector)}: the frame-level representations are first processed into contextual features, then aggregated at multiple temporal resolutions, and finally attentively pooled into a fixed-dimensional vector that summarizes the entire utterance;
(3) \textbf{Contrastive Fingerprint Training (Vector to Fingerprint)}: the vector is optimized to be robust to benign operations, and sensitive to tampering attacks using contrastive learning;
(4) \textbf{Binary Fingerprint Encoding (Fingerprint to Bit)}: the trained fingerprint is discretized into a binary representation;  
(5) \textbf{Segment-Wise Watermarking (Bit to Watermark)}: the binary fingerprint is embedded into the original audio through segment-wise watermarking, making the fingerprint self-contained.

The speech integrity verification in SpeechVerifier independently performs two parallel paths on the published audio: (1) regenerating the fingerprint via the same extraction pipeline, and (2) extracting the embedded watermark via the watermark decoding process. The two resulting binary codes are then compared using Hamming distance to determine whether the speech has been attacked.

\subsection{Fingerprint Generation and Watermarking}\label{Self-Contained Fingerprint Generation}
\textbf{Step 1. Frame-Level Feature Encoding (Speech to Representation)}
We utilize the pre-trained wav2vec 2.0 model~\cite{baevski2020wav2vec} to extract frame-level representations from the original audio before publishing. This step serves as a necessary preprocessing stage for fingerprint generation. It converts continuous waveform signals into structured sequences of frame-level representations that preserve essential acoustic information. These representations have demonstrated effectiveness in downstream tasks such as automatic speech recognition~\cite{baevski2021unsupervised} and speaker verification~\cite{fan2021exploring}. Formally, the feature encoder $\varepsilon : \mathcal{X} \to \mathcal{Z}$ maps raw audio waveforms $\mathcal{X}$ to a sequence of latent representations $\mathbf{z}_1, \mathbf{z}_2, \dots, \mathbf{z}_T$,  where each $\mathbf{z}_t \in \mathbb{R}^{d_{z}}$ denotes the frame-level acoustic feature at time $t$, and $T$ is the total number of output frames.

\textbf{Step 2. Multiscale Acoustic Feature Extraction and Summarization (Representation to Vector).}
Given the frame-level representations $\mathbf{z}_1, \mathbf{z}_2, \dots, \mathbf{z}_T$ obtained from Step 1, this step constructs a fixed-dimensional vector that summarizes the speech across different temporal granularities. The multiscale feature extractor $\mathcal{F}$ consists of two components: (a) a bidirectional long short-term memory (BiLSTM) network that transforms the input frame-level representations into contextual hidden states, and (b) a multiscale pooling operation that averages the hidden states over phoneme-, word-, and phrase-level windows (size 20, 50, and 100, respectively)
, producing a sequence of multiscale features $\mathbf{h}_1, \mathbf{h}_2, \dots, \mathbf{h}_K$ (see Appendix~\ref{apx:A3}).

To summarize these features into an utterance-level vector, we apply self-attentive pooling~\cite{lin2017structured}. This mechanism assigns higher weights to more informative components, with attention weight computed as: $w_n = \frac{\exp(\phi(h_n))}{\sum_{t=1}^{K} \exp(\phi(h_t))}, \quad \text{where }\phi(\cdot)$ is a feedforward network. The weighted sum yields a fixed-dimension vector: $\mathbf{v}' = \sum_{n=1}^{K} w_n \cdot h_n$, which is referred to as the utterance-level vector. To obtain a more compact representation for fingerprint optimization, a projection module is applied to reduce the dimensionality of $\mathbf{v}'$, yielding the final fingerprint vector $\mathbf{v} \in \mathbb{R}^{d_{v}}$.

\textbf{Step 3. Contrastive Fingerprint Training (Vector to Fingerprint).}
Given the fixed-length vector $\mathbf{v}$ obtained from Step 2, we optimize it to serve as a distinctive audio fingerprint that is robust to benign operations and sensitive to malicious tampering attacks. To this end, we adopt contrastive learning~\cite{oord2018representation} to guide the training of all preceding modules. During the training, a batch of original speech samples is randomly selected, where each sample serves as an anchor. For each anchor, we generate: \textbf{Positive pairs}, consisting of the anchor and its benign variants (e.g., compression and re-encoding), and \textbf{Negative pairs}, consisting of the anchor and its tampered variants (e.g., substitution and deletion). Detailed operations are listed in Appendix~\ref{apx:A2}. The contrastive loss is defined as 
\begin{equation}
\mathcal{L}_c = -\frac{1}{B} \sum_{i=1}^{B} \frac{1}{P} \sum_{j=1}^{P} 
\log \frac{ 
\exp\left( \tilde{\mathbf{v}}_{i}^{\text{Orig.}\top} \tilde{\mathbf{v}}_{i,j}^{\text{Benign}} / \tau \right) 
}
{
\sum\limits_{\substack{k=1,\ k \neq i}}^{N} 
\exp\left( {\tilde{\mathbf{v}}_{i}^{\text{Orig.}\top}} \tilde{\mathbf{v}}_{i,k} / \tau \right)
},
\end{equation}
where $B$ is the number of anchors in the batch, $P$ is the number of benign variants per anchor, and $N$ denotes the total number of comparison samples for each anchor, including its own benign and tampered variants as well as embeddings from other anchors in the batch. $\tau$ is the temperature parameter. $\tilde{\mathbf{v}}_{i}^{\text{Orig.}}$ denotes the L2-normalized embedding of the $i$-th anchor, $\tilde{\mathbf{v}}_{i,j}^{\text{Benign}}$ denotes the embedding of its $j$-th benign variant, and $\tilde{\mathbf{v}}_{i,k}$ enumerates all embeddings in the batch, including benign, tampered and unrelated samples. 

This contrastive learning above encourages the model to bring the anchor closer to its benign variants while pushing it away from tampered and unrelated samples in the embedding space. As a result, the fixed-length vector is optimized to serve as a distinctive audio fingerprint that is robust to benign operations while remaining sensitive to malicious tampering attacks.

\textbf{Step 4. Binary Fingerprint Encoding (Fingerprint to Bit).}
Compared to full-precision vectors in continuous space, binary representations are more suitable for compact storage (e.g., embedding into a watermark) and fast retrieval (e.g., through bit-wise comparison). Therefore, we convert the continuous fingerprint vector $\mathbf{v} \in \mathbb{R}^{d_v}$ into a binary code $\mathbf{b} \in \{-1, +1\}^d$. This binary fingerprint can directly be embedded into the audio signal. Specifically, we apply a \texttt{tanh} activation followed by a \texttt{sign} function at the final projection layer to obtain the binarized output. As demonstrated later in the evaluation, the binarized fingerprint preserves its discriminative characteristics of $\mathbf{v}$, i.e., robust to benign operations while sensitive to tampering attacks.

\textbf{Step 5. Segment-Wise Watermarking (Bit to Watermark).}
To enable self-contained verification, we embed the binary fingerprint directly into the speech signal. Inspired by AudioSeal~\cite{roman2024proactive}, we aim to develop a high-capacity and inaudible audio watermarking method based on the Encodec framework. While AudioSeal targets short watermarks for copyright-protection (i.e., 16 bits), SpeechVerifier must embed much longer fingerprints (e.g., 256 bits) to support speech integrity verification. To meet this requirement, we propose a segment-wise watermarking scheme. Given an input waveform $\mathcal{X}$ of duration $T$ seconds and its binary fingerprint $\mathbf{b}$, both are divided into $N$ non-overlapping segments:
\begin{equation}
\mathcal{X} = [\mathcal{X}^{(1)}, \dots, \mathcal{X}^{(N)}] \quad\text{and}\quad \mathbf{b} = [\mathbf{b}^{(1)}, \dots, \mathbf{b}^{(N)}],
\end{equation}
where each $\mathcal{X}^{(n)}$ spans $T/N$ seconds and each $\mathbf{b}^{(n)}$ contains $d/N$ bits. For each audio segment $\mathcal{X}^{(n)}$, we embed $\mathbf{b}^{(n)}$ into the Encodec embedding space and generate a watermark signal $\delta^{(n)}$. The watermarked segment is then formed as: $\tilde{\mathcal{X}}^{(n)} = \mathcal{X}^{(n)}+\delta^{(n)}$. Finally, the watermarked segments $[\tilde{\mathcal{X}}^{(1)}, \dots, \tilde{\mathcal{X}}^{(N)}]$ are concatenated, yielding the final self-verifiable audio. Notably, since watermark only incurs subtle perturbations and remains imperceptible to human listeners, the embedded binary fingerprint can be reliably extracted from the watermarked speech audio $\tilde{\mathcal{X}}$ without degradation.

\subsection{Dual-Path Speech Integrity Verification\label{sec:verification}}
SpeechVerifier employs a dual-path mechanism to assess the integrity of the published speech $\tilde{\mathcal{X}}$:

\textbf{Path A: Fingerprint Generation from Published Speech.} The published speech audio is processed using the same fingerprint generation pipeline described in Section~\ref{Self-Contained Fingerprint Generation}. The fingerprint $\mathbf{b}'$  is computed as $\mathbf{b}' = \text{sign}(\mathcal{F}(\varepsilon(\tilde{\mathcal{X}})))$, where $\varepsilon$ and $\mathcal{F}$ denote the feature encoder and multiscale extractor, respectively, and $\text{sign}(\cdot)$ denotes the final binarization function.

\textbf{Path B: Watermark Extraction.} The published speech audio $\tilde{\mathcal{X}}$ is processed in the inverse manner of Step 5 to decode the embedded watermark (i.e., the original binary fingerprint). From each segment $\tilde{\mathcal{X}}^{(n)}$, we extract the bit chunk $\hat{\mathbf{b}}^{(n)}$ using the watermark decoder, and then reconstruct the full watermark as $\hat{\mathbf{b}} = [\hat{\mathbf{b}}^{(1)}, \hat{\mathbf{b}}^{(2)}, \dots, \hat{\mathbf{b}}^{(N)}]$.

Finally, the integrity of the published audio is verified by comparing the generated fingerprint $\mathbf{b}'$ with the extracted watermark $\hat{\mathbf{b}}$. This is done by computing the Hamming distance as follows.
\vspace{-1mm}
\[
d_H(\mathbf{b}', \hat{\mathbf{b}}) \leq \theta \quad \Rightarrow \quad \text{Accept}; \quad \text{otherwise} \quad \text{Reject},
\]
\vspace{-2mm}
where $\theta$ is a decision threshold set based on the development data from public datasets.

\vspace{-2mm}
\section{Experiments}
\subsection{Experiment Setup}
\textbf{Dataset.} To train and evaluate the performance of the proposed SpeechVerifier, we use \textbf{VoxCeleb1}~\cite{nagrani2017voxceleb} dataset, which includes over 150,000 utterances from 1,251 celebrities. These audio samples are collected from interviews and public videos, providing the conditions that reflect real-world speech recordings. In addition, we use the test subset from \textbf{LibriSpeech}~\cite{panayotov2015librispeech} dataset. LibriSpeech is a corpus of approximately 1,000 hours of English read speech, sourced from public domain audiobooks. This setup allows us to perform a cross-domain evaluation to assess model generalization. More details about the datasets and the preprocessing steps are provided in Appendix~\ref{apx:B2}.

\textbf{Implementation details.} \textbf{Fingerprint model:} We use the pre-trained Wav2Vec2.0 Base model as the acoustic feature extractor, obtained from the official repository\footnote{https://github.com/facebookresearch/fairseq/tree/main/examples/wav2vec}. A two-layer BiLSTM with a hidden size of 512 follows the feature extractor. Multiscale pooling is used with window sizes of 20, 50, and 100 frames with a stride of 10 frames. A two-layer projection head then maps features into a 256-dimensional vector. \textbf{Watermark model:} The pre-trained AudioSeal model\footnote{https://github.com/facebookresearch/audioseal} is used to embed and extract fingerprints as watermark payloads. To improve the watermarking capacity, we divide both the carrier audio and the fingerprint into 16 segments. Each segment carries a 16-bit watermark, leading to a total payload of 256 bits per audio sample. \textbf{Train:} We exploit benign and malicious operations (see Appendix~\ref{apx:A2}) and the original audio samples for contrastive learning, with temperature set as 0.05. A cosine annealing learning rate schedule is used, gradually decreasing the learning rate from $1\times10^{-3}$ to $1\times10^{-5}$ over the training.

\textbf{Evaluation Metrics.} The threshold in Section~\ref{sec:verification} is determined on the development dataset ($\theta=42$). We compare the dual-path bit error against $\theta$ for binary classification, where benign and malicious operations are treated as positive and negative samples, respectively. Evaluation metrics include true positive rate (TPR), false positive rate (FPR), true negative rate (TNR), false negative rate (FNR), equal error rate (EER), and area under the curve (AUC). Additionally, we present data visualizations, as well as cosine similarity and Hamming distance analyses, to demonstrate effectiveness.

\begin{table}[h]
    \centering
    \caption{Results of benign operation (positive) acceptance on VoxCeleb and LibriSpeech.}
    \resizebox{0.95\textwidth}{!}{
    \begin{tabular}{lcccc|cccc|cc}
        \toprule
        \multirow{2}{*}{\textbf{Operation}} &
        \multicolumn{4}{c|}{\textbf{VoxCeleb}} &
        \multicolumn{4}{c|}{\textbf{LibriSpeech}} &
        \multirow{2}{*}{\textbf{Semantic}} &
        \multirow{2}{*}{\textbf{Identity}} \\
        \cmidrule(lr){2-5} \cmidrule(lr){6-9}
        & \textbf{TPR} & \textbf{FPR} & \textbf{AUC} & \textbf{EER}
        & \textbf{TPR} & \textbf{FPR} & \textbf{AUC} & \textbf{EER} \\
        \midrule
        Compression         & 1.00 & 0.00 & 1.00 & 0.00 & 1.00 & 0.03 & 1.00 & 0.01 & \checkmark & \checkmark \\
        Reencoding            & 1.00 & 0.00 & 1.00 & 0.00 & 1.00 & 0.02 & 1.00 & 0.01 & \checkmark & \checkmark \\
        Resampling          & 1.00 & 0.00 & 1.00 & 0.00 & 0.97 & 0.01 & 1.00 & 0.02 & \checkmark & \checkmark \\
        Noise suppression   & 1.00 & 0.00 & 1.00 & 0.00 & 0.98 & 0.01 & 1.00 & 0.01 & \checkmark & \checkmark \\
        \midrule
        \textbf{Overall}    & 1.00 & 0.00 & 1.00 & 0.00 & 0.99 & 0.02 & 1.00 & 0.02 & - & - \\
        \bottomrule
    \end{tabular}
    }
    \label{tab:benign-performance}
\end{table}

\subsection{Results}
\textbf{Robustness to benign operations.}
Table~\ref{tab:benign-performance} presents the performance of the proposed SpeechVerifier in accepting published speech samples subjected to benign audio operations, as defined in Appendix~\ref{apx:A2}. We focus on evaluating how well SpeechVerifier accepts positive samples with harmless modifications (TPR) and whether it mistakenly accepts maliciously tampered speech (FPR). The numbers of positive and negative test samples are balanced in Table~\ref{tab:benign-performance}. When trained and evaluated on different subsets of VoxCeleb, SpeechVerifier achieves 100\% TPR and 0\% FPR across \textit{all} listed benign operations, demonstrating strong robustness to non-malicious transformations. For cross-dataset evaluation on LibriSpeech, using a model trained on VoxCeleb, the overall TPR/FPR slightly change to 99\% and 2\%, respectively, indicating good generalizability across datasets.

\textbf{Sensitivity to tampering attacks.}
Table~\ref{tab:malicious-performance} evaluates the ability of \textbf{SpeechVerifier} to reject malicious tampering attacks that alter the semantic content or speaker identity. Specifically, we consider tampering attacks including deletion, splicing, silencing, substitution, reordering, as well as deepfake-based manipulations such as text-to-speech synthesis (TTS) and voice conversion, as detailed in Appendix~\ref{apx:A2}. In this evaluation, tampering operations (actual negatives) to the published audio are expected to be rejected with a high true negative rate, while minimizing the false negative rate, which reflects incorrect rejection of benign samples. Notably, SpeechVerifier achieves 100\% and 98\% overall TNR, and 0\% and 1\% overall FNR on the VoxCeleb (in-domain) and LibriSpeech (cross-domain) datasets, respectively. These results highlight SpeechVerifier's strong sensitivity to tampering attacks. A more detailed breakdown by tampering strength (e.g., minor, moderate and severe) is provided in Appendix~\ref{apx:B5}.

\vspace{-5mm}
\begin{table}[t]
    \centering
    \caption{Results of malicious operation (negative) rejection on VoxCeleb and LibriSpeech.}
    \resizebox{0.95\textwidth}{!}{
    \begin{tabular}{lcccc|cccc|cc}
        \toprule
        \multirow{2}{*}{\textbf{Operation}} &
        \multicolumn{4}{c|}{\textbf{VoxCeleb}} &
        \multicolumn{4}{c|}{\textbf{LibriSpeech}} &
        \multirow{2}{*}{\textbf{Semantic}} &
        \multirow{2}{*}{\textbf{Identity}} \\
        \cmidrule(lr){2-5} \cmidrule(lr){6-9}
        & \textbf{TNR} & \textbf{FNR} & \textbf{AUC} & \textbf{EER}
        & \textbf{TNR} & \textbf{FNR} & \textbf{AUC} & \textbf{EER} \\
        \midrule
        Deletion     & 1.00 & 0.00 & 1.00 & 0.00 & 0.99 & 0.00 & 1.00 & 0.00 & \xmark & \checkmark \\
        Splicing    & 1.00 & 0.00 & 1.00 & 0.00 & 0.99 & 0.08 & 0.99 & 0.01 & \xmark & \checkmark \\
        Silencing         & 1.00 & 0.00 & 1.00 & 0.00 & 0.98 & 0.00 & 1.00 & 0.01 & \xmark & \checkmark \\
        Substitution      & 1.00 & 0.00 & 1.00 & 0.00 & 0.99 & 0.00 & 1.00 & 0.01 & \xmark & \checkmark \\
        Reordering           & 1.00 & 0.00 & 1.00 & 0.00 & 0.98 & 0.00 & 1.00 & 0.00 & \xmark & \checkmark \\
        Text-to-speech       & 1.00 & 0.00 & 1.00 & 0.00 & 1.00 & 0.00 & 1.00 & 0.00 & \xmark & \checkmark \\  
        Voice conversion     & 1.00 & 0.00 & 1.00 & 0.00 & 0.98 & 0.00 & 1.00 & 0.01 & \checkmark & \xmark \\ 
        \midrule
        \textbf{Overall}    & 1.00 & 0.00 & 1.00 & 0.00 & 0.98 & 0.01 & 1.00 & 0.02 & - & - \\
        \bottomrule
    \end{tabular}}
    \label{tab:malicious-performance}
\end{table}

\subsection{Evaluation of Multiscale Feature and Binarized Fingerprints}
\begin{figure}[t]
  \centering
  \begin{minipage}[b]{0.67\textwidth}
    \centering
    \begin{subfigure}[b]{0.49\linewidth}
      \centering
      \includegraphics[width=\linewidth]{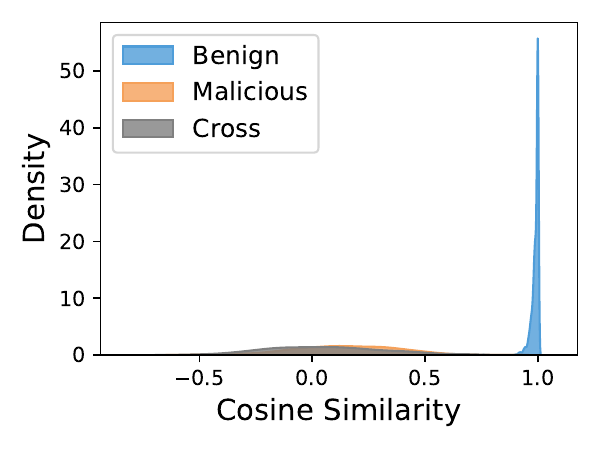}
      \caption{}
      \label{fig:sim_ours}
    \end{subfigure}
    \hfill
    \begin{subfigure}[b]{0.49\linewidth}
      \centering
      \includegraphics[width=\linewidth]{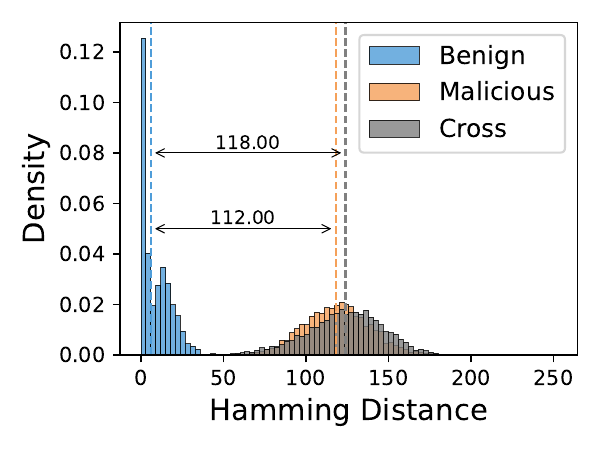}
      \caption{}
      \label{fig:hamm_ours}
    \end{subfigure}
      \caption{(a) Extracted feature similarity; (b) binarized fingerprint Hamming distance.}
      \label{fig:dist_ours}
    \end{minipage}
  \hfill
  \begin{minipage}[b]{0.32\textwidth}
    \centering
    \begin{subfigure}[b]{\linewidth}
      \centering
      \includegraphics[width=\linewidth]{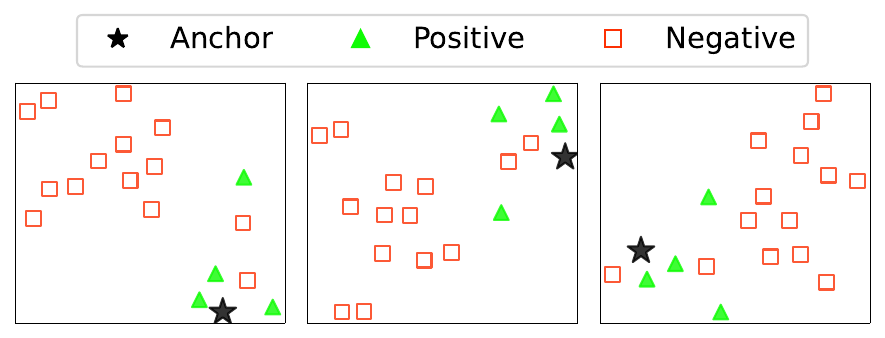}
      \caption{}
      \label{fig:tsne_before}
    \end{subfigure}   

    \begin{subfigure}[b]{\linewidth}
      \centering
      \includegraphics[width=\linewidth]{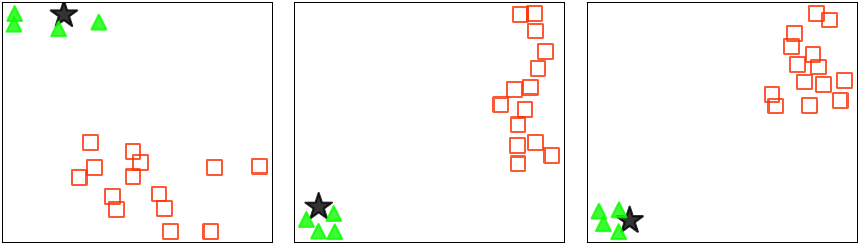}
      \caption{}
      \label{fig:tsne_after}
    \end{subfigure}
    
    \caption{t-SNE visualizations of speech samples: (a) before training; (b) after training.}
    \label{fig:tsne_both}
  \end{minipage}
\end{figure}

Figure~\ref{fig:dist_ours} presents two types of analysis: (a) cosine similarity between extracted multiscale features and (b) Hamming distance between binarized fingerprints. In Figure~\ref{fig:sim_ours}, benign-original pairs show high similarity values close to 1.0, while malicious-original and cross-original pairs exhibit significantly lower similarity scores, indicating that the learned multiscale features effectively capture the differences between benign operations and malicious tampering attacks. Here, ``cross'' refers to the arbitrarily selected unrelated audio samples. In Figure~\ref{fig:hamm_ours}, binarized fingerprints of benign processed samples yield low Hamming distances from their corresponding retrieved watermarks, whereas malicious and cross pairs have much larger distances, with a clear separation gap of nearly 112-118 bits. This indicates that the binarization process preserves discriminability and can differentiate tampering attacks from benign operations by using a simple threshold.

Figure~\ref{fig:tsne_both} illustrates the t-SNE visualizations of the extracted multiscale features before and after training. Specifically, Figure~\ref{fig:tsne_before} and Figure~\ref{fig:tsne_after} show the relative distribution of anchors (original speech), positives (after benign operations), and negatives (after malicious operations) in the latent space. Before training, anchor and positive samples are scattered and overlap with negatives, indicating that the initial features extracted are not well separated. After training, anchors and positives form tight clusters, while negatives are clearly separated. This suggests that contrastive learning-trained multiscale feature extraction effectively learns embeddings that distinguish benign operations from malicious tampering, which explains the strong performance of SpeechVerifier. Additional visualization evidences are provided in Appendix~\ref{apx:B8}.

\subsection{Comparison with Other Deepfake Detection Methods}
We finally evaluate SpeechVerifier as a deepfake detector\footnote{To avoid confusion, SpeechVerifier is used here for deepfake detection, where ``positive” now refers to deepfake samples to be identified.}. Specifically, we compare SpeechVerifier with state-of-the-art methods including RawNet2~\cite{tak2021end} and AASIST~\cite{jung2022aasist}, two end-to-end models developed for the ASVspoof challenge~\cite{yamagishi2021asvspoof}, and widely used for audio spoofing detection. We utilize a zero-shot TTS model YourTTS~\cite{casanova2022yourtts} to generate deepfake speech segments, and substitute them for varying proportions (10\%, 25\%, 50\%, 75\% and 90\%) of the original speech. Next, the deepfake samples are mixed with an equal amount of clean speech to ensure fair evaluation.

From Table~\ref{tab:method-comparison}, both RawNet2 and AASIST perform best given the highest substitution ratio at 90\%, achieving up to 83\% TPR by RawNet2. However, when decreasing the substitution ratio, RawNet2 and AASIST both show significant degradation regarding the ability of detecting deepfake substitution samples. For example, at the ratio of 10\%, the AUC of RawNet2 drops to 65\% and EER increases to 38\%, indicating diminished sensitivity to subtle spoofing. Similarly, AASIST exhibits similar performance degradation under the same condition. In contrast, SpeechVerifier constantly achieves very good detection performance across all substitution levels (TPR=1.00, FPR=0.00, AUC=1.00, EER=0.00), demonstrating the superiority of the proposed proactive defense design.

Since synthetic deepfake audio lacks embedded watermarks, the fingerprint-watermark verification process becomes essentially random, making tampering attacks easy to detect. Even minor substitutions alter the extracted fingerprint and disrupt the embedded watermark simultaneously, resulting in a mismatch and enabling reliable detection of tampering. Further analysis of each module's contributions is provided in the ablation studies in Appendix~\ref{apx:B7}.

\vspace{-4mm}
\begin{table}[H]
    \centering
    \caption{Performance of Deepfake detection at varying substitution ratios}
    \resizebox{\textwidth}{!}{
    \begin{tabular}{lcccc|cccc|cccc}
        \toprule
        \textbf{Deepfake Ratio} &
        \multicolumn{4}{c|}{\textbf{RawNet2}} &
        \multicolumn{4}{c|}{\textbf{AASIST}} &
        \multicolumn{4}{c}{\textbf{SpeechVerifier (ours)}} \\
        \cmidrule(lr){2-5} \cmidrule(lr){6-9} \cmidrule(lr){10-13}
        & \textbf{TPR} & \textbf{FPR} & \textbf{AUC} & \textbf{EER}
        & \textbf{TPR} & \textbf{FPR} & \textbf{AUC} & \textbf{EER}
        & \textbf{TPR} & \textbf{FPR} & \textbf{AUC} & \textbf{EER} \\
        \midrule
        Substitute 10\%   & 0.58 & 0.34 & 0.65 & 0.38 & 0.39 & 0.17 & 0.64 & 0.39 & \textbf{1.00} & \textbf{0.00} & \textbf{1.00} & \textbf{0.00} \\
       Substitute 25\%   & 0.62 & 0.34 & 0.68 & 0.37 & 0.37 & 0.17 & 0.64 & 0.42 & \textbf{1.00} & \textbf{0.00} & \textbf{1.00} & \textbf{0.00} \\
        Substitute 50\%   & 0.59 & 0.34 & 0.65 & 0.38 & 0.44 & 0.17 & 0.67 & 0.39 & \textbf{1.00} & \textbf{0.00} & \textbf{1.00} & \textbf{0.00} \\
        Substitute 75\%   & 0.68 & 0.34 & 0.72 & 0.34 & 0.54 & 0.17 & 0.75 & 0.31 & \textbf{1.00} & \textbf{0.00} & \textbf{1.00} & \textbf{0.00} \\
        Substitute 90\%  & 0.83 & 0.34 & 0.81 & 0.27 & 0.62 & 0.17 & 0.76 & 0.33 & \textbf{1.00} & \textbf{0.00} & \textbf{1.00} & \textbf{0.00} \\
        \bottomrule
    \end{tabular}}
    \label{tab:method-comparison}
\end{table}
\vspace{-8mm}

\section{Conclusion}
In this paper, we have proposed SpeechVerifier, a proactive speech integrity verification design. SpeechVerifier employs multiscale feature extraction and contrastive learning to generate robust acoustic fingerprints. Then, it embeds the generated fingerprints into audio signals through watermarking. The fingerprint is designed to be sensitive to malicious tampering attacks but robust to benign operations, thus not affecting the normal audio distribution and usage. This approach enables the general public to conveniently verify the published speech audio's integrity in a self-contained manner, i.e., by extracting and comparing embedded watermark messages with corresponding acoustic fingerprints. Extensive experiments demonstrate that SpeechVerifier achieves very good detection performance under various tampering attack scenarios while staying robust to benign operations.

\clearpage
\appendix
\section{SpeechVerifier Design and Operation Definitions}
\subsection{Overall Algorithm}

\begin{algorithm}[H]
\caption{SpeechVerifier Training and Deployment}
\begin{algorithmic}[1]
\State \textbf{Input:} Raw speech $\mathcal{X}$, benign operations $\mathcal{T}_b(\cdot)$, malicious operations $\mathcal{T}_m(\cdot)$, Wav2Vec2.0 encoder $\varepsilon$, multiscale feature extractor $\mathcal{F}$
\State \textbf{Output:} Watermarked speech $\tilde{\mathcal{X}}$
\For{$e = 1, 2, \ldots, \text{epochs}$}
    \For{$b=1,2,\ldots, \text{batches}$}
        \State $\mathcal{X}^{\text{benign}} \gets \mathcal{T}_b(\mathcal{X})$, \quad $\mathcal{X}^{\text{malicious}} \gets \mathcal{T}_m(\mathcal{X})$ \Comment{Generate positive and negative variants}
        
        \State \textbf{Step 1: Frame-level feature extraction}
        \State $\mathcal{Z} \gets \varepsilon(\mathcal{X})$ \Comment{Extract frame-level acoustic representations}

        \State \textbf{Step 2: Multiscale feature summarization}
        \State $\mathbf{h}_n \gets \mathcal{F}(\mathcal{Z})$ \Comment{BiLSTM and multiscale pooling}
        \For{$n = 1, \ldots, K$}
            \State $w_n \gets \frac{\exp(\phi(\mathbf{h}_n))}{\sum_{t=1}^{K}\exp(\phi(\mathbf{h}_t))}$ \Comment{Attentive weight computation}
        \EndFor
        \State $\mathbf{v}' \gets \sum_{n=1}^{K}w_n \cdot \mathbf{h}_n$ \Comment{Utterance-level representation}
        \State $\mathbf{v} \gets \text{Proj}(\mathbf{v}')$ \Comment{Final fingerprint vector}
        
        \State \textbf{Step 3: Contrastive fingerprint training}
        \State Compute contrastive loss $\mathcal{L}_{\text{c}}$
        \State Update $\mathcal{F}$, $\phi$, $\text{Proj}$ via backpropagation
    \EndFor
\EndFor

\State \textbf{Step 4: Binary fingerprint encoding}
\State $\mathbf{b} \gets \text{sign}(\tanh(\text{Proj}(\text{AttPool}(\mathcal{F}(\varepsilon(\tilde{X}))))))$ \Comment{Extract binary fingerprint}

\State \textbf{Step 5: Segment-wise watermarking}
\State Split $\mathcal{X}$ and $\mathbf{b}$ into $N$ segments: $[\mathcal{X}^{(1)}, \dots, \mathcal{X}^{(N)}]$, $[\mathbf{b}^{(1)}, \dots, \mathbf{b}^{(N)}]$
\For{$n = 1, \dots, N$}
    \State $\delta^{(n)} \gets \text{WatermarkEmbedder}(\mathcal{X}^{(n)}, \mathbf{b}^{(n)})$ 
    \State $\tilde{\mathcal{X}}^{(n)} \gets \mathcal{X}^{(n)} + \delta^{(n)}$ \Comment{Embed watermark into audio}
\EndFor
\State $\tilde{\mathcal{X}} \gets \text{Concat}(\tilde{\mathcal{X}^{(1)}}, \dots, \tilde{\mathcal{X}}^{(N)})$ \Comment{Get watermarked audio} 
\end{algorithmic}
\end{algorithm}

\begin{algorithm}[H]
\caption{SpeechVerifier Verification}
\begin{algorithmic}[1]
\State \textbf{Input:} Published speech $\tilde{\mathcal{X}}$, wav2vec2.0 encoder $\varepsilon$, trained multiscale feature extractor $\mathcal{F}$, projection module $\text{Proj}$, attentive pooling $\text{AttPool}$, $\text{WatermarkExtractor}$
\State \textbf{Output:} Verification result (Accept or Reject)
\vspace{0.5em}
\State \textbf{Path A: Fingerprint extraction}
\State $\mathbf{b}' \gets \text{sign}(\tanh(\text{Proj}(\text{AttPool}(\mathcal{F}(\varepsilon(\tilde{\mathcal{X}}))))))$
\vspace{0.5em}
\State \textbf{Path B: Segment-wise watermark extraction}
\State Split $\tilde{\mathcal{X}}$ into $N$ segments: $\tilde{\mathcal{X}}^{(1)}, \dots, \tilde{\mathcal{X}}^{(N)}$
\For{$n = 1$ to $N$}
    \State $\hat{\mathbf{b}}^{(n)} \gets \text{WatermarkExtractor}(\tilde{\mathcal{X}}^{(n)})$
\EndFor
\State $\hat{\mathbf{b}} \gets \text{Concat}(\hat{\mathbf{b}}^{(1)}, \dots, \hat{\mathbf{b}}^{(N)})$
\vspace{0.5em}
\State \textbf{Integrity decision}
\If{$d_H(\mathbf{b}', \hat{\mathbf{b}}) \leq \theta$}
    \State \Return Accept
\Else
    \State \Return Reject
\EndIf
\end{algorithmic}
\end{algorithm}
 
\subsection{Definition of Benign Operations and Malicious Tampering}\label{apx:A2}
We simulate two categories of audio modifications: \textit{benign operations} and \textit{malicious tampering}. Benign operations refer to legitimate processing steps encountered during audio storage, transmission, or distribution. These operations do not change the semantic content or the speaker identity of the speech. In contrast, malicious tampering refers to intentional alterations designed to distort either the semantic meaning or the identity of the speaker. We detail each operation below and summarize their characteristics in Table~\ref{tab:audio_ops}.

\begin{table}[h]
\centering
\caption{Summary of audio operations.}
\label{tab:audio_ops}
\begin{tabular}{@{}p{3.2cm}p{5.0cm}p{4.8cm}@{}}
\toprule
\textbf{Operation} & \textbf{Example} & \textbf{Implementation} \\
\midrule
\multicolumn{3}{@{}l@{}}{\textbf{\textit{Benign Operations}}} \\
Compression & Podcasts, news broadcasts, online meetings & \texttt{ffmpeg -i in.wav -b:a 128k tmp.mp3; ffmpeg -i tmp.mp3 out.wav} \\
Reencoding & Saving or uploading audio files & \texttt{ffmpeg -i in.wav out.wav} \\
Resampling & Low-bandwidth communication & \texttt{torchaudio.transforms.Resample} \\
Noise Suppression & Social media platforms & \texttt{RMS-based frame muting} \\
\midrule
\multicolumn{3}{@{}l@{}}{\textbf{\textit{Malicious Operations}}} \\
Deletion & Removing ``not'' in ``I do not agree'' & \texttt{VAD + remove voiced portion} \\
Splicing & Inserting ``not'' into ``I do agree'' & \texttt{Insert voiced segment} \\
Substitution & Replacing ``agree'' with ``disagree'' & \texttt{Swap waveform segment} \\
Silencing & Muting ``not'' in ``I do not agree'' & \texttt{Mute VAD-detected region} \\
Reordering & Changing sentence order & \texttt{Segment + shuffle + concat} \\
Voice Conversion & Changing timbre (speaker identity) & \texttt{torchaudio.sox\_effects} \\
Text-to-Speech & Generate new speech with speaker's timbre & \texttt{YourTTS (zero-shot synthesis)} \\
\bottomrule
\end{tabular}
\end{table}

\paragraph{Compression.} Lossy compression is applied by converting the waveform to MP3 at 128 kbps and decoding it back to WAV. This simulates typical processing in podcasts and streaming platforms. We use FFmpeg: \texttt{ffmpeg -i input.wav -b:a 128k temp.mp3; ffmpeg -i temp.mp3 output.wav}.

\paragraph{Reencoding.} The waveform is re-encoded to 16-bit PCM WAV format without compression. This simulates storage or uploading scenarios where minor numerical alterations may occur. Implemented with: \texttt{ffmpeg -i input.wav output.wav}.

\paragraph{Resampling.} Audio is downsampled (e.g., from 16 kHz to 8 kHz) and then upsampled back, simulating low-bandwidth or legacy systems. Implemented with: \texttt{torchaudio.transforms.Resample}. 

\paragraph{Noise Suppression.} To simulate automatic noise suppression utilized by social media and streaming platforms, the waveform is divided into overlapping frames. Frames with low root-mean-square (RMS) energy are muted.

\paragraph{Deletion.} A portion of speech (not silence) is removed from the speech. For example, deleting ``not'' from ``I do not agree'' changes the meaning entirely.

\paragraph{Splicing.} A short segment of speech from the same speaker is spliced into the waveform. For example, inserting ``not'' into the phrase ``I do agree'' reverses its original semantic meaning. 

\paragraph{Substitution.} A segment of speech is replaced with another waveform segment of equal length from the same speaker. For instance, replacing ``agree'' with ``disagree'' fundamentally changes the intended meaning.

\paragraph{Silencing.} A portion of speech (not silence or noise) is deliberately muted by setting its amplitude to zero. For instance, muting the word ``not'' in ``I do not agree'' leads to a reversed interpretation.

\paragraph{Reordering.} The speech is segmented, rearranged, and concatenated to change the semantic content. For instance, reordering ``I never said she stole my money'' into ``She stole my money, I never said'' distorts the original meaning and can lead to an opposite interpretation.

\paragraph{Voice Conversion.} The speech is manipulated to alter its speaker identity while preserving the linguistic content. We simulate this by shifting pitch (e.g., +4 semitones) using SoX effects. This modification can make the utterance sound like it was spoken by someone else. We implement this using \texttt{torchaudio.sox\_effects.apply\_effects\_tensor}.

\paragraph{Text-to-Speech.} We synthesize speech from text using a pre-trained text-to-speech (TTS) model, YourTTS\footnote{\url{https://github.com/Edresson/YourTTS}}~\cite{casanova2022yourtts}, which supports multilingual and zero-shot speaker adaptation. This attack can generate speech that closely mimics the speaker's voice with arbitrary semantic content.

\paragraph{Different Levels of Tampering.}
To evaluate the performance under varying conditions, we define three levels of tampering: \textit{Minor}, \textit{Moderate}, and \textit{Severe}. Specifically, at the Minor level, tampering operations --- including deletion, splicing, silencing, and substitution --- alter about 10\% of the original audio content (alteration ratio = 0.1). At the Moderate level, these same operations alter 30\% of the audio (alteration ratio = 0.3). At the Severe level, 50\% of the audio is altered (alteration ratio = 0.5), and this level also includes reordering operations, which disrupts the logical structure of the speech.

\subsection{Explanation of Malicious Tampering over Different Granularities}\label{apx:A3}
Table~\ref{tab:tampering_examples} presents representative examples of malicious tampering at the phoneme, word, and phrase levels. These examples illustrate how manipulations at different temporal granularities can alter the meaning of speech. They also motivate the use of multiscale pooling with window sizes of 20, 50, and 100 frames, which are designed to capture such variations in real-world scenarios.
\begin{table}[h]
\centering
\caption{Examples of malicious tampering at different levels of granularity}
\begin{tabular}{l p{5cm} p{6cm}}
\toprule
\textbf{Granularity} & \textbf{Example} & \textbf{Description} \\ 
\midrule
Phoneme-level & 
Change ``bed'' to ``bad'' (English); change ``mā'' (mother) to ``mǎ'' (horse) (Mandarin) & 
Altering a single phoneme or syllable can lead to subtle yet meaningful changes. These edits are often difficult to detect but can reverse or distort the intended meaning. \\

Word-level & 
Insert ``not'' into ``He is guilty'' to form ``He is not guilty''; replace ``approved'' with ``denied'' & 
Tampering at the word level through insertion, deletion, or substitution can directly modify semantic content, leading to misleading interpretations. \\

Phrase-level & 
Change ``Negotiations will begin immediately'' to ``Negotiations will be delayed indefinitely'' & 
Reordering or replacing entire phrases can fabricate new narratives while maintaining natural-sounding speech, making the tampering more deceptive. \\
\bottomrule
\end{tabular}
\label{tab:tampering_examples}
\end{table}
\clearpage

\section{Experimental Setup and Extended Results}
\subsection{Implementation Details}\label{apx:B1}
To supplement Section 5.1, we provide a detailed description of the model architecture and training configuration. 

\textbf{Model.}
We use the pretrained Wav2Vec2.0 Base model~\footnote{\url{https://github.com/facebookresearch/fairseq/blob/main/examples/wav2vec}} to extract 768-dimensional frame-level acoustic features. These are passed to a two-layer Bidirectional LSTM (BiLSTM) with an input size of 768, a hidden size of 512 (i.e., 256 per direction), and a dropout rate of 0.25. To capture temporal features at multiple resolutions, we apply average pooling with window sizes of 20, 50, and 100 frames, with a stride of 10 frames, implemented using \texttt{avg\_pool1d} along the time axis. The pooled outputs are aggregated by an attentive pooling module consisting of a linear-tanh-linear projection. The resulting weighted sum forms the utterance-level embedding, followed by dropout with a rate of 0.2. This embedding is fed into a two-layer MLP projection head with dimensions $768 \rightarrow 512 \rightarrow 256$, with ReLU activation between layers. The final output vector is L2-normalized and passed through a $\tanh$ function to constrain values to the range $[-1, 1]$, yielding the continuous-valued fingerprint. For segment-wise watermarking, we use the pretrained AudioSeal model\footnote{\url{https://github.com/facebookresearch/audioseal}} to embed and extract binary fingerprints as watermarks. Each audio is divided into 16 non-overlapping segments, with each segment embedded with a 16-bit binary watermark, resulting in a total payload size of 256 bits per audio.

\textbf{Training.}
SpeechVerifier is trained using a cosine annealing learning rate schedule, decaying from $1 \times 10^{-3}$ to $1 \times 10^{-5}$ over 50 epochs. The contrastive loss is temperature-scaled with $\tau = 0.05$. Training is conducted on 4 NVIDIA Quadro RTX 8000 GPUs using distributed data parallelism (DDP).

\subsection{Dataset and Evaluation Details}\label{apx:B2}
We use two public speech datasets: \textbf{VoxCeleb} and \textbf{LibriSpeech}. For VoxCeleb, the development set is used for training and the test set for evaluation. For LibriSpeech, we use only the \texttt{test-clean} subset for evaluation. All audio files are converted to WAV format and resampled to 16 kHz. 

\textbf{Preprocessing.}
We randomly sample 5{,}000 utterances from the VoxCeleb development set for model training. For evaluation, we sample 500 utterances each from the VoxCeleb test set and the LibriSpeech \texttt{test-clean} subset. To stabilize the training and ensure data quality, we retain only utterances with durations between 2 and 20 seconds.

For each valid utterance, we generate two sets of augmented variants for contrastive learning:

\textit{Benign Augmentations:}  
These are modifications that preserve both speaker identity and semantic content. The details can be found in Appendix~\ref{apx:A2}.

\textit{Malicious Augmentations:}  
These include tampering operations intended to alter speaker identity and semantic content. The details can be found in Appendix~\ref{apx:A2}.

\textbf{Evaluation Metric}
For evaluation, we consider benign and malicious as positive and negative classes, respectively. TP is the number of benign samples correctly classified, and FN is the number of benign samples incorrectly classified as malicious. FP is the number of malicious samples incorrectly classified as benign, and TN is the number of malicious samples correctly rejected. The following metrics are computed:

\begin{itemize}
\item True Positive Rate (TPR):
\[
\text{TPR} = \frac{\text{TP}}{\text{TP} + \text{FN}},
\]

\item False Positive Rate (FPR):
\[
\text{FPR} = \frac{\text{FP}}{\text{FP} + \text{TN}},
\]

\item True Negative Rate (TNR):
\[
\text{TNR} = \frac{\text{TN}}{\text{TN} + \text{FP}},
\]

\item False Negative Rate (FNR):
\[
\text{FNR} = \frac{\text{FN}}{\text{FN} + \text{TP}}
\]

\item Equal Error Rate (EER): The error rate at the decision threshold where FPR = FNR.

\item Area Under the ROC Curve (AUC): Computed by integrating the Receiver Operating Characteristic (ROC) curve over various thresholds.
\end{itemize}

\subsection{Similarity Distribution using MFCC Feature}\label{apx:B3}
\begin{figure}[h]
    \centering
    \resizebox{0.7\textwidth}{!}{
    \begin{tabular}{cc}
        \includegraphics[width=0.20\textwidth]{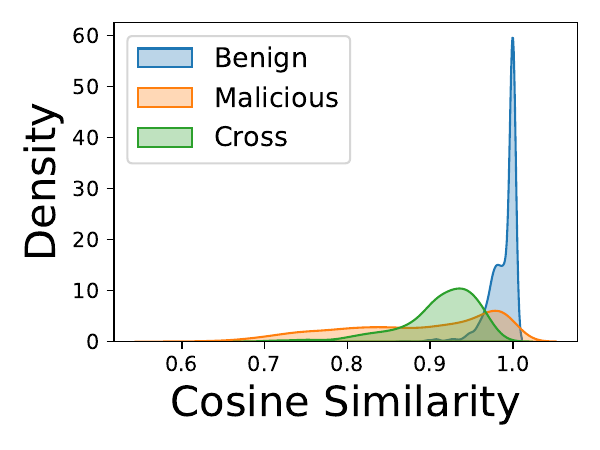} &
        \includegraphics[width=0.20\textwidth]{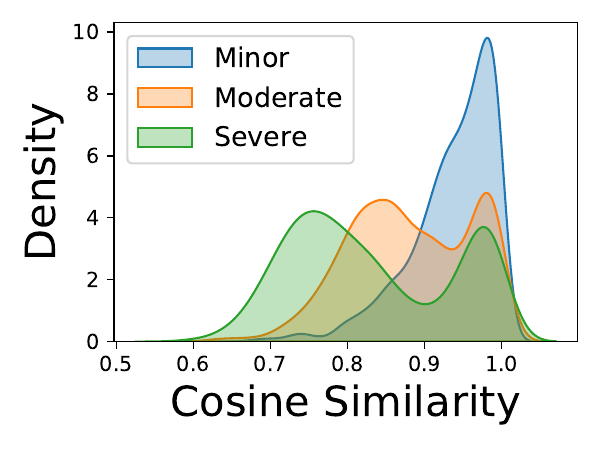}
    \end{tabular}
    }
    \caption{Probability distributions: (a) MFCC embedding similarity to the original audio under different modifications; (b) MFCC embedding similarity to the original audio at different tampering levels.}
    \label{fig:mfcc_observation}
\end{figure}

To complement the observations in Section~3.3, we present similarity distributions computed using handcrafted Mel-frequency cepstral coefficients (MFCC) instead of wav2vec2 embeddings. As shown in Figure~\ref{fig:mfcc_observation}, the similarity distributions between original and modified audio samples using MFCC features exhibit trends similar to those observed with wav2vec2-based representations. Specifically, the distributions corresponding to benign and malicious modifications overlap, and the similarity scores tend to decrease as the extent of tampering increases. This indicates that MFCC-based similarity comparison can only measure the extent of modification but does not effectively distinguish between different types of modifications.
\clearpage
\subsection{Evaluation on Semantic and Identity Changes under Benign and Malicious Operations}\label{apx:B4}
We evaluate the impact of different audio modifications on both semantic integrity and speaker identity consistency. Semantic preservation is quantified using word error rate (WER) computed from a pre-trained automatic speech recognition (ASR) model\footnote{\url{https://github.com/facebookresearch/fairseq/blob/main/examples/wav2vec}}, \texttt{facebook/wav2vec2-base-960h}, a CTC-based ASR model. Speaker identity preservation is measured by cosine similarity between embeddings extracted using the pre-trained speaker verification (SV) model\footnote{\url{https://huggingface.co/speechbrain/spkrec-ecapa-voxceleb}}, \texttt{speechbrain/spkrec-ecapa-voxceleb}.

As shown in Table~\ref{tab:wer_identity}, benign operations (e.g., compression, recording, resampling, noise suppression) result in low WER ($\le$8.24) and high identity similarity ($\ge$0.78), indicating that they largely preserve both semantic content and speaker identity. 

In contrast, malicious operations introduce substantial degradation. WER increases steadily with the severity of deletion, splicing, silencing, and substitution, reflecting significant semantic changes. These operations, however, generally maintain high identity similarity because they retain the original timbre. Notably, voice conversion results in relatively low WER, but significantly reduces identity similarity (41.60\%), since it deliberately alters the speaker's timbre. 

To further investigate the nonzero WER observed under benign operations, we manually examined the ASR outputs. Most transcription errors were minor substitutions or alignment shifts that did not affect the overall meaning. This suggests that the observed WER in these cases reflects limitations of the ASR model and metric sensitivity rather than genuine semantic distortion.

\begin{table}[h]
\centering
\caption{WER and Identity Similarity under Different Operations}
\begin{tabular}{lcc}
\toprule
\textbf{Operation} & \textbf{WER \%} & \textbf{Identity Similarity \%} \\
\midrule
\multicolumn{3}{l}{\textbf{\textit{Benign Operations}}} \\
Compression         & 1.15  & 95.60 \\
Recoding            & 0.26  & 99.99 \\
Resampling          & 6.87  & 78.00 \\
Noise suppression   & 8.24  & 94.52 \\
\midrule
\multicolumn{3}{l}{\textbf{\textit{Malicious Operations}}} \\
Deletion (minor)    & 21.65 & 99.04 \\
Deletion (moderate) & 40.20 & 96.84 \\
Deletion (severe)   & 62.32 & 93.29 \\
Splicing (minor)      & 31.24 & 99.00 \\
Splicing (moderate)   & 52.45 & 97.35 \\
Splicing (severe)     & 78.36 & 96.55 \\
Silencing (minor)     & 30.76 & 98.16 \\
Silencing (moderate)  & 53.79 & 90.13 \\
Silencing (severe)    & 75.74 & 75.96 \\
Substitution (minor)     & 23.22 & 98.33 \\
Substitution (moderate)     & 48.12 & 94.36 \\
Substitution (severe)     & 63.03 & 90.72 \\
Reordering             & 69.55 & 99.53 \\
Text-to-speech      &   -   &   -  \\
Voice conversion    &  8.60 & \textbf{41.60} \\
\bottomrule
\end{tabular}
\label{tab:wer_identity}
\end{table}

\clearpage

{\subsection{Results of Fine-Grained Malicious Operations Rejection}\label{apx:B5}

\begin{table}[h]
    \centering
    \caption{Results of fine-grained malicious operation rejection on VoxCeleb and LibriSpeech.}
    \resizebox{\textwidth}{!}{
    \begin{tabular}{lcccc|cccc|cc}
        \toprule
        \multirow{2}{*}{\textbf{Operation}} &
        \multicolumn{4}{c|}{\textbf{VoxCeleb}} &
        \multicolumn{4}{c|}{\textbf{LibriSpeech}} &
        \multirow{2}{*}{\textbf{Semantic}} &
        \multirow{2}{*}{\textbf{Identity}} \\
        \cmidrule(lr){2-5} \cmidrule(lr){6-9}
        & \textbf{TNR} & \textbf{FNR} & \textbf{AUC} & \textbf{EER}
        & \textbf{TNR} & \textbf{FNR} & \textbf{AUC} & \textbf{EER} \\
        \midrule
        Deletion (minor)     & 1.00 & 0.00 & 1.00 & 0.00 & 0.98 & 0.00 & 1.00 & 0.00 & \xmark & \checkmark \\
        Deletion (moderate)  & 1.00 & 0.00 & 1.00 & 0.00 & 0.99 & 0.00 & 1.00 & 0.00 & \xmark & \checkmark \\
        Deletion (severe)    & 1.00 & 0.00 & 1.00 & 0.00 & 0.99 & 0.00 & 1.00 & 0.00 & \xmark & \checkmark \\
        Splicing (minor)    & 1.00 & 0.00 & 1.00 & 0.00 & 0.99 & 0.16 & 0.98 & 0.06 & \xmark & \checkmark \\
        Splicing (moderate) & 1.00 & 0.00 & 1.00 & 0.00 & 0.99 & 0.07 & 0.99 & 0.03 & \xmark & \checkmark \\
        Splicing (severe)   & 1.00 & 0.01 & 1.00 & 0.01 & 0.99 & 0.01 & 1.00 & 0.01 & \xmark & \checkmark \\
        Silencing (minor)         & 1.00 & 0.00 & 1.00 & 0.00 & 0.99 & 0.00 & 1.00 & 0.00 & \xmark & \checkmark \\
        Silencing (moderate)      & 1.00 & 0.00 & 1.00 & 0.00 & 0.98 & 0.00 & 1.00 & 0.00 & \xmark & \checkmark \\
        Silencing (severe)        & 1.00 & 0.00 & 1.00 & 0.00 & 0.98 & 0.00 & 1.00 & 0.02 & \xmark & \checkmark \\
        Substitution (minor)      & 1.00 & 0.00 & 1.00 & 0.00 & 0.99 & 0.00 & 1.00 & 0.01 & \xmark & \checkmark \\
        Substitution (moderate)   & 1.00 & 0.00 & 1.00 & 0.00 & 0.99 & 0.00 & 1.00 & 0.01 & \xmark & \checkmark \\
        Substitution (severe)     & 0.99 & 0.00 & 1.00 & 0.00 & 0.99 & 0.00 & 1.00 & 0.00 & \xmark & \checkmark \\
        Reordering           & 1.00 & 0.00 & 1.00 & 0.00 & 0.98 & 0.00 & 1.00 & 0.00 & \xmark & \checkmark \\
        Text-to-speech       & 1.00 & 0.00 & 1.00 & 0.00 & 1.00 & 0.00 & 1.00 & 0.00 & \xmark & \checkmark \\  
        Voice conversion     & 1.00 & 0.00 & 1.00 & 0.00 & 0.98 & 0.00 & 1.00 & 0.01 & \checkmark & \xmark \\ 
        \midrule
        \textbf{Overall}    & 1.00 & 0.00 & 1.00 & 0.00 & 0.99 & 0.02 & 1.00 & 0.01 & - & - \\
        \bottomrule
    \end{tabular}}
    \label{tab:malicious-finegrained-performance}
\end{table}

\subsection{Watermarked Speech Quality}
\begin{table}[h]
    \centering
    \caption{Audio quality metrics.}
    \begin{tabular}{lcccc}
        \toprule
        \textbf{Methods} & \textbf{SI-SNR} & \textbf{PESQ} & \textbf{STOI} & \textbf{LSD} \\
        \midrule
        SpeechVerifier   & 25.14 & 4.28 & 0.998 & 0.111 \\
        \bottomrule
    \end{tabular}
    \label{tab:watermark-metrics}
\end{table}

We evaluate the perceptual quality of watermarked speech using four objective metrics. (1)Scale-Invariant Signal to Noise Ratio (SI-SNR) quantifies waveform-level distortion in decibels (dB). Higher values indicate less distortion. (2)Perceptual Evaluation of Speech Quality (PESQ)~\cite{rix2001perceptual} ranges from 1.0 (poor) to 4.5 (excellent), and reflects perceived speech quality. (3)Short-Time Objective Intelligibility (STOI)~\cite{taal2010short} ranges from 0 to 1, with higher values indicating better intelligibility. (4)Log Spectral Distance (LSD) measures spectral deviation between original and watermarked speech, lower values indicate greater spectral fidelity.

As shown in Table~\ref{tab:watermark-metrics}, our proposed SpeechVerifier has little perceptual degradation. The high SI-SNR and PESQ scores, along with near-perfect intelligibility (STOI) and low spectral error (LSD), demonstrate that the watermarking process preserves both fidelity and intelligibility, making it suitable for practical deployment.

\subsection{Ablation Studies}\label{apx:B7}
SpeechVerifier consists of four core modules: an acoustic feature encoder, a multiscale feature extractor, an attentive pooling module, and a contrastive learning objective. To assess the contribution of each module, we conduct ablation studies by replacing each module with alternatives and evaluating the performance.

For the feature encoder, we compare handcrafted Mel-Frequency Cepstral Coefficients (MFCC) with deep speech representations obtained from a pretrained wav2vec model. To evaluate the impact of the multiscale feature extractor, we remove the multiscale pooling, directly use the output of BiLSTM. For temporal pooling, we substitute attentive pooling with average pooling. Regarding the loss function, we compare InfoNCE with Triplet Loss, which are widely used in contrastive learning.

As shown in Table~\ref{tab:ablation-speechverifier}, each module contributes to the overall performance. Replacing the wav2vec encoder with handcrafted MFCC features results in only a slight performance drop, suggesting that MFCC can also serve as a lightweight alternative. Removing the multiscale feature extractor leads to a significant degradation, indicating the importance of extracting both global and local temporal patterns. Substituting attentive pooling with average pooling reduces performance, indicating that the attention mechanism provides better frame selection for embedding generation. Finally, replacing InfoNCE with Triplet Loss results in a substantial performance decline, demonstrating that InfoNCE is more effective for learning discriminative embeddings in our task.

\begin{table}[ht]
    \centering
    \caption{Ablation studies on acoustic feature encoder, feature extractor, temporal pooling scheme, and loss function.}
    \begin{tabular}{l|ccccc}
        \toprule
        \textbf{Method Variant} & \textbf{TPR} & \textbf{FPR} & \textbf{AUC} & \textbf{EER} \\
        \midrule
        SpeechVerifier (wav2vec $\rightarrow$ MFCC)        & 0.9979 & 0.0021 & 0.9999 & 0.0021 \\
        SpeechVerifier (Multiscale $\rightarrow$ w/o Multiscale)  & 0.7443 & 0.1224 & 0.8847 & 0.1958 \\
        SpeechVerifier (AttentivePooling $\rightarrow$ AvgPooling)      & 0.9875 & 0.0163 & 0.9988 & 0.0144 \\
        SpeechVerifier (InfoNCEloss $\rightarrow$ TripletLoss)     & 0.8594 & 0.0146 & 0.9577 & 0.1073 \\
        \textbf{SpeechVerifier}     & \textbf{0.9979} & \textbf{0.0016} & \textbf{1.0000} & \textbf{0.0013} \\
        \bottomrule
    \end{tabular}
    \label{tab:ablation-speechverifier}
\end{table}

\subsection{Visualization of Multiscale Feature Extraction}\label{apx:B8}

\begin{figure}[h]
    \centering
    \begin{subfigure}[b]{0.9\linewidth}
        \centering
        \includegraphics[width=\linewidth]{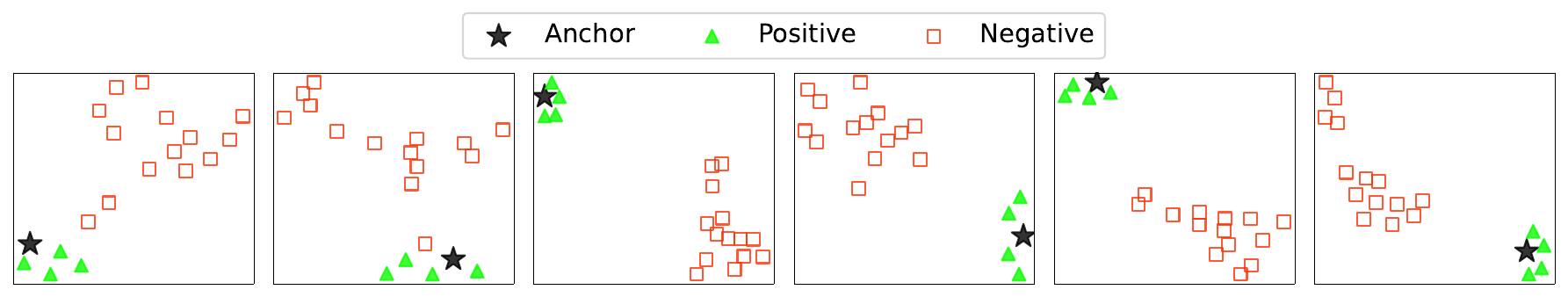}
        \label{fig:tsne_1}
    \end{subfigure}
    \begin{subfigure}[b]{0.9\linewidth}
        \centering
        \includegraphics[width=\linewidth]{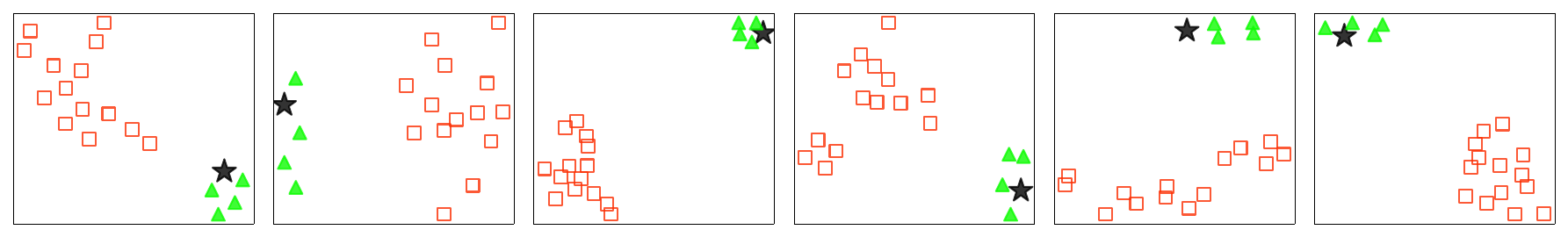}
        \label{fig:tsne_2}
    \end{subfigure}
    \begin{subfigure}[b]{0.9\linewidth}
        \centering
        \includegraphics[width=\linewidth]{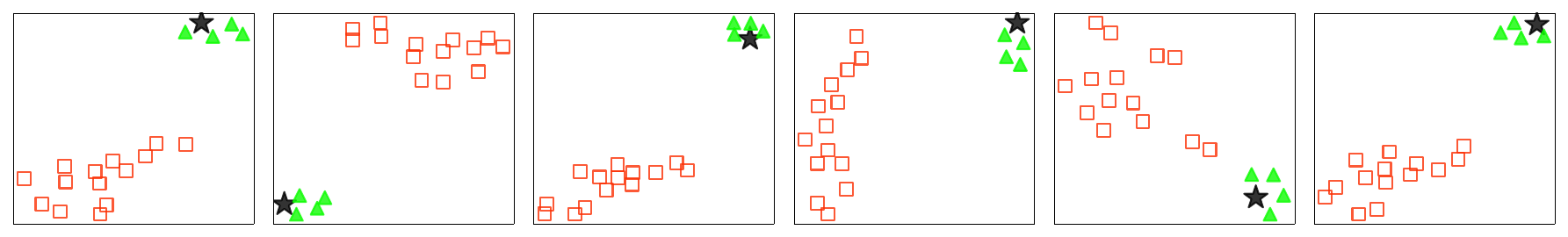}
        \label{fig:tsne_3}
    \end{subfigure}
    \begin{subfigure}[b]{0.9\linewidth}
        \centering
        \includegraphics[width=\linewidth]{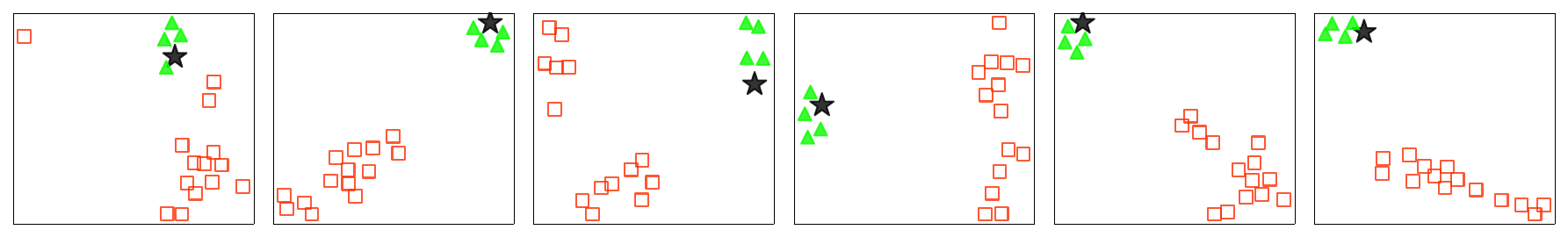}
        \label{fig:tsne_4}
    \end{subfigure}    
    \caption{t-SNE visualizations of speech samples (after training).}
    \label{fig:tsne_all}
\end{figure}

\clearpage
\section{Limitations and Broader Impact}
\subsection{Limitations}
While SpeechVerifier demonstrates strong robustness to benign audio operations and effective detection of malicious tampering, there are several aspects that can be further improved:
\begin{itemize}
\item SpeechVerifier currently identifies whether an audio has been tampered with, but does not localize the exact region of tampering. Future work could explore providing localized detection for more precise forensic analysis.

\item Although we have performed cross-dataset evaluation using LibriSpeech, our experiments mainly focus on English speech under relatively clean conditions. The generalization of SpeechVerifier to noisy, multilingual, and more diverse real-world tampering operations needs further investigation.

\item SpeechVerifier assumes that the input audio is sufficiently long ($\ge$ 2s) to support segment-wise watermark embedding and extraction. For very short audio clips, the process of embedding and extracting watermarks may become unreliable. Moreover, such short clips often lack meaningful semantic content and offer limited value for tampering, making protection less critical in practice. However, to extend the applicability of SpeechVerifier, future work is needed to explore more reliable verification methods for short speech segments. 
\end{itemize}

\subsection{Broader Impact}
SpeechVerifier aims to proactively protect the integrity of publicly shared speech content. By detecting audio tampering and impersonation attacks, it can effectively mitigate the spread of misinformation, safeguard individual reputations, and uphold public trust in digital media. Furthermore, since SpeechVerifier operates without relying on external references or original recordings, it substantially reduces verification costs and enhances scalability. 

% {\small
% \bibliographystyle{plainnat}
% \bibliography{ref}
% }

\end{document}